\documentclass[12pt,aip,showpacs,a4paper,prb]{revtex4}
\usepackage{amsmath}

\begin{document}

\hyphenation{res-pec-ti-ve-ly}
\hyphenation{e-le-ment}
\hyphenation{co-rres-ponds}
\hyphenation{coin-ci-des}
\title{SPINOR EXTENDED LORENTZ-FORCE-LIKE EQUATIONS AS CONSEQUENCE OF A 
SPINORIAL STRUCTURE OF SPACE-TIME}

\author{J. Buitrago} 
\author{S. Hajjawi}
\affiliation{Department of Astrophysics of the
University of La Laguna,    
Avenida Francisco Sanchez, s/n, 38205, La Laguna, Tenerife, Spain} 
\email{jgb@ll.iac.es}
\date{May 2006, revised November 2006}

\begin{abstract}
As formerly shown by one of us and briefly discussed in the Introduction below, 
the special relativistic dynamical equation of the Lorentz force type
can be regarded as a consequence of a succession of space-time
dependent infinitesimal Lorentz boosts and rotations. This
insight indicates that the Lorentz-Force-like equation has a
 fundamental meaning in physics.  In this paper we show how
this result may be spinorially obtained starting out from the application
of an infinitesimal element of $SL(2,C)$ to the individual spinors, which
are here regarded as being more fundamental objects than four-vectors.
In this way we get a set of new dynamical spinor equations {\em inducing}
the extended Lorentz-Force-like equation in the Minkowski space-time and
geometrically obtain the spinor form of the electromagnetic field-tensor.
The term extended refers to the dynamics of some additional degrees of
freedom that may be associated with the intrinsic spin, namely with the
dynamics of three space-like mutually orthogonal four-vectors, all of
them orthogonal to the linear four-momentum of the object under
consideration which finally, in the particle's proper frame, are identified with 
the generators of $SU(2)$.
\end{abstract}

\maketitle

\section{INTRODUCTION}

Some years ago one of us in \cite{buitrago} found
that the Lorentz force equation may be regarded as a consequence of the
infinitesimal transformation, within the Lorentz Group, which relates the 
four-momentum
of the particle at (proper) times $\tau$ and $\tau +
\delta\tau$ respectively. This result is obtained from purely
geometrical considerations so one can conclude that, at the classical
level, any force acting on a particle is restricted by the geometry of
space-time to act in a very special, not arbitrary at all, way.
Henceforth we shall call this fairly general result
\footnote{The Geometrical Principle is, in fact, not only applicable
to the four-momenta, but to any four-vector representing any physical
magnitude.} the $Geometrical$ $Principle$.
Now, the motivation for the next section is that of re-interpreting
this Principle as a consequence of the geometry of
the $Spinor$ $Space$
$\mathcal{S}(2,\mathcal{C})$, in order to emphasize the idea that spinors are 
more
fundamental objects than four-vectors and may thus be regarded as
the elementary building blocks from which physical four-vectors are
made of. We recall that any element $Q$ of $SL(2,C)$ acting on
$\mathcal{S}(2,\mathcal{C})$, univocally defines
a (proper) Lorentz transformation and reassume this idea by saying that
the spinor spaces associated to any two distinct
inertial observers are related through
an element of $SL(2,C)$,  
which in turn
induces a linear transformation on $\overline{\mathcal{S}(2,\mathcal{C})}$
through
the element $\overline{Q}$ of $\overline{SL(2,C)}$ \footnote{This is better
understood if we recall that
$\mathcal{M}_{4}$ is a sub-space of the four-complex-dimensional
vector-space $\mathcal{S}
(2,\mathcal{C}) \bigotimes \overline{\mathcal{S}(2,\mathcal{C})}$.}.
In \cite{buitrago} one starts out by considering that the four-momentum
of a particle, which currently follows a trajectory in space-time,
at proper times $\tau$ and $\tau + \delta\tau$ respectively, should be
related through $\delta\Lambda^{\alpha}_{\quad \beta}$ by:\\
\begin{displaymath}
p^{\alpha}(\tau+\delta\tau)=\delta\Lambda^{\alpha}_{\quad \beta}p^{\beta}
(\tau)
\end{displaymath}
where $\delta\Lambda^{\alpha}_{\quad \beta}$ is an infinitesimal element
of the proper Lorentz group of transformations. It can be easily shown
that  this equation leads to a Lorentz-force-like
equation and therefore, the Lorentz-Force is no more a semi-empirical
result, but  finds its theoretic justification within this context.
%
%
%
% SECTION SPINORIAL FORMULATION
%
%
%
\section{SPINORIAL FORMULATION OF THE GEOMETRICAL PRINCIPLE}
As stated in the Introduction above, the components of any spinor associated 
with a particle
following a trajectory in space-time at (proper) times $\tau$ and $\tau
+\delta\tau$ respectively must be related through an infinitesimal element
$\delta t_{B}^{\quad A}$ of $SL(2,C)$, so the key equation is (we adopt
the standard spinor algebra notation as given in \cite{prrw})

\begin{equation} \label{1.1}
\xi^{A}(\tau + \delta\tau)=\delta t_{B}^{\quad A}\xi^{B},
\end{equation}
where $\delta t_{B}^{\quad A}$ is of the form
\begin{displaymath}
\delta t_{B}^{\quad A}= \left[ e^{\frac{1}{2}(\delta\vec{\omega}+
i\delta\vec{\theta})\vec{\sigma}}\right] \equiv
\left( e^{\delta\vec{\alpha}\vec{\sigma}}\right)_{B}^{\quad A},
\end{displaymath}
being $\delta\vec{\omega}=(\delta\omega_{1},\delta\omega_{2},
\delta\omega_{3})$, $\delta\vec{\theta}=(\delta\theta_{1},\delta\theta_{2},
\delta\theta_{3})$ the infinitesimal parameters related to boosts and
rotations respectively and $\vec{\sigma}=(\sigma_{1},\sigma_{2},
\sigma_{3})$ the vector containing the three Pauli spin matrices.\\
This starting point should lead us, via the $Geometrical$ $Principle$, to
the equations of motion of the particle in their spinorial form, as well
as to a spinorial representation of the force-field acting on it, if
our assertions are true. Since $\delta t_{B}^{\quad A}$ is infinitesimal,
we expand the exponential term and keep only first-order-terms, so
(\ref{1.1}) becomes
\begin{equation} \label{1.2}
\xi^{A}(\tau + \delta\tau)=\xi^{B}\delta t_{B}^{\quad A}=
\xi^{A}(\tau) + \left(\delta\vec{\alpha} \cdot \vec{\sigma}\right)
_{B}^{\quad A}
\xi^{B}(\tau) ,
\end{equation}
being $(\delta\vec{\alpha} \cdot \vec{\sigma})_{B}^{\quad A}$
the matrix operator
\begin{equation} \label{1.3}
(\delta\vec{\alpha} \cdot \vec{\sigma})_{B}^{\quad A}=
\left( \begin{array}{cc}
\delta\alpha_{3} & \delta\alpha_{1}+i\delta\alpha_{2} \\
\delta\alpha_{1}-i\delta\alpha_{2} & -\delta\alpha_{3} \\
\end{array} \right)
\end{equation}
which acts on the spinor $\xi^{B}$ transforming it infinitesimally. On the
other side, if a change in velocity occurs, it should be
proportional to the force-field acting on the particle and the lapse of
proper time so we next define two fields $\vec{\epsilon}(x)
,\vec{\beta}(x)$
in such a way that
\begin{displaymath}
\begin{array}{c}
\delta\vec{\omega} = \mathcal{K} \vec{\epsilon}(\bar{x})\delta\tau, \\
\delta\vec{\theta}= \mathcal{K} \vec{\beta}(\bar{x})\delta\tau,\\
\end{array}
\end{displaymath}
being $\mathcal{K}$ some constant.
In turn (\ref{1.2}) can be manipulated to give
\begin{equation} \label{1.4}
\begin{array}{c}
\delta\xi^{A} \equiv \xi^{A}(\tau+\delta\tau) -\xi^{A}(\tau) \\
= \xi^{B}(\tau)(\delta\vec{\alpha} \cdot \vec{\sigma})_{B}^{\quad A} \\
=\epsilon^{BC}\xi_{C}(\tau)(\delta\vec{\alpha} \cdot \vec{\sigma})_{B}^
{\quad A} \\
=-\xi_{C}(\tau)\epsilon^{CB}(\delta\vec{\alpha} \cdot \vec{\sigma})_
{B}^{\quad A}= \mathcal{K}\delta\tau\xi_{C}(\tau)\phi^{CA}= \mathcal{K}
\delta\tau\phi^{AC}\xi_{C}(\tau) . \\
\end{array}
\end{equation}
In this way we have introduced the symmetric form $\phi^{AB}=
-\epsilon^{AC}(\delta\vec{\alpha} \cdot \vec{\sigma})_{C}^{\quad B}$.
From (\ref{1.4}) one
obtains
\begin{equation} \label{1.5}
\frac{d\xi^{A}}{d\tau} \equiv \dot{\xi}^{A}=\mathcal{K}\phi^{AB}\xi_{B},
\end{equation}
where $\phi^{AB}$ is explicitely given by
\begin{equation} \label{1.6}
\phi^{AB}= \frac{1}{2}
\left( \begin{array}{cc}
-\left[ \epsilon_{1}+\beta_{2}\right] + i\left[\epsilon_{2}-\beta_{1}
\right] & \epsilon_{3}+i\beta_{3} \\
\epsilon_{3}+i\beta_{3} & \left[\epsilon_{1}-\beta_{2}\right]+i
\left[\beta_{1}+\epsilon_{2}\right] \\
\end{array}\right) .
\end{equation}
Suppose now a massive particle currently acted on by an applied external
field and let us see which equation of motion (\ref{1.5}) leads to.
Since the particle has non-null mass $m$, the spinorial representation
of its four-momentum must be written as superposition of two null directions
in the form
\begin{displaymath}
p^{AA'}=\pi^{A}\bar{\pi}^{A'} + \eta^{A}\bar{\eta}^{A'},
\end{displaymath}
resctricted by the condition
\begin{equation} \label{1.7}
p^{AA'}p_{AA'}=2\vert\pi^{A}\eta_{A}\vert^{2}=m^{2}.
\end{equation}
{ } \\
Now, taking the derivative with respect to proper time and using
(\ref{1.5}) we get (we have set $\mathcal{K}=1$ for short)
\begin{equation} \label{1.8}
\begin{array}{c}
\dot{p}^{AA'}=\dot{\pi}^{A}\bar{\pi}^{A'}+\pi^{A}\dot{\bar{\pi}}^{A'}+
\dot{\eta}^{A}\bar{\eta}^{A'}+\eta^{A}\dot{\bar{\eta}}^{A'} \\
=\phi^{AB}\pi_{B}\bar{\pi}^{A'}+\bar{\phi}^{A'B'}\bar{\pi}_{B'}\pi^{A}+
\phi^{AB}\eta_{B}\bar{\eta}^{A'}+\bar{\phi}^{A'B'}\bar{\eta}_{B'}\eta^{A} \\
=\epsilon^{A'B'}\phi^{AB}\pi_{B}\bar{\pi}_{B'}+
\epsilon^{AB}\bar{\phi}^{A'B'}\pi_{B}\bar{\pi}_{B'} +
\epsilon^{A'B'}\phi^{AB}\eta_{B}\bar{\eta}_{B'} +
\epsilon^{AB}\bar{\phi}^{A'B'}\eta_{B}\bar{\eta}_{B'} \\
=\left( \epsilon^{AB}\bar{\phi}^{A'B'}  + \epsilon^{A'B'}\phi^{AB} \right)
p_{BB'} \\
=F^{AA'BB'}p_{BB'},
\end{array}
\end{equation}
where
\begin{equation} \label{1.9}
\epsilon^{AB}\bar{\phi}^{A'B'}+\epsilon^{A'B'}\phi^{AB}=F^{AA'BB'}.
\end{equation}
Spinorially, the general equation of motion is then given by,
for $\mathcal{K}=1$ \\
\begin{equation} \label{1.10}
\dot{p}^{AA'}= F^{AA'BB'}p_{BB'},
\end{equation}
so the index-lowered expression becomes
\begin{equation} \label{1.11}
\dot{p}_{AA'}= F_{AA'BB'}p^{BB'}.
\end{equation}\\
In the particular case of an external electromagnetic field, by
replacing
\begin{displaymath}
\begin{array}{c}
\vec{\epsilon}=\vec{E}, \\
{ } \\
\vec{\beta}=\vec{B}, \\
\end{array}
\end{displaymath}
and $\mathcal{K}=q/m$
we find out that (\ref{1.10})  is just the spinorial representation of the
Lorentz Force. Note, however, that in this way, the components of $F_{AA'
BB'}$ are obtained from those of $\phi_{AB}$ not just as a spinorial
transcription of the electromagnetic tensor, as it is usually done
\cite{Stewart}. Instead, they have been here deduced from the geometric
standpoint emphasized in (\ref{1.1}), our starting point. On the other
hand, in (\ref{1.1}) we could have considered any other dynamical spinors
instead of those associated to the four-momentum of a massive particle,
so this result  manifest how a spinorial structure hidden behind
the minkowskian geometry of space-time at local level, makes the dynamical
evolution of any physical four-vector to adopt the form of a Lorentz-Force-
like equation. \\
In the case of an electromagnetic field and from (\ref{1.6}), the components
of $\phi^{AB}$ are related to the physical components of the field by
\begin{equation} \label{1.12}
\begin{array}{c}
\phi^{00}= -\frac{1}{2}\left[(E_ {x}+B_{y})-i(E_{y}-B_{x})\right], \\
{ } \\
\phi^{01}=\phi^{10}=\frac{1}{2}(E_{z}+i B_{z}), \\
{ } \\
\phi^{11}=\frac{1}{2}\left[(E_{x}-B_{y})+i(E_{y}+B_{x})\right], \\
\end{array}
\end{equation} \\
and the one-spinor
equation of motion is re-writtten as \\
\begin{equation} \label{1.13}
\dot{\xi}^{A}= \frac{q}{m}\phi^{AB}\xi_{B}.
\end{equation} \\
For the general case (no need to specify the nature of the force acting
on the particle), we have \\
\begin{equation} \label{1.14}
\dot{\xi}^{A}=\mathcal{K}\phi^{AB}\xi_{B},
\end{equation}\\
being therefore the (spinorial) equation of motion for the four-momentum
\begin{equation} \label{1.15}
\dot{p}^{AA'}=\mathcal{K}F^{AA'BB'}p_{BB'},
\end{equation}
where $\mathcal{K}$ becomes a
contant factor depending on some intrinsic feature
of the particle and $\phi^{AB}$ is the spinorial representation of the
force-field under consideration. Finally, the equivalent standard-tensorial
(index-lowered) expression is
\begin{equation} \label{1.16}
\dot{p}_{\alpha}=\mathcal{K}F_{\alpha\beta}p^{\beta}
\end{equation}
with $F_{\alpha\beta}$ necessary obeying to
\begin{equation} \label{1.17}
F_{\alpha\beta}=-F_{\beta\alpha} .
\end{equation}
The last requirement is set by the (local) geometry of space-time and
therefore the trajectory of a particle under the action of any force-field
is restricted by the condition (\ref{1.17}). Again, we remark that this
idea constitutes the Geometrical Principle. The fact that $F_{\alpha
\beta}$ must be anti-symmetric is a consequence of the minkowskian
geometry, which in turn, in this new context, arises from a spinorial structure
hidden behind it. On the other side, not only the form of the Lorentz-Force
equation, as it is done, may be deduced: in the particular case that
$F_{\alpha\beta}$ may be expressed in terms of a vectorial field $A_
{\alpha}(x)$ as
\begin{displaymath}
F_{\alpha\beta}=\partial_{\alpha}A_{\beta}-\partial_{\beta}A_{\alpha},
\end{displaymath}
Maxwell equations
for free fields are an immediate consequence.
%
%
% SECTION SPINOR EQUATIONS EQUIVALENT TO THE EXTENDED LORENTZ FORCE-LIKE 
EQUATION
\section{SPINOR EQUATIONS EQUIVALENT TO THE EXTENDED LORENTZ FORCE-LIKE 
EQUATION}
As we have already seen from (\ref{1.5}) and the set of previous relations
to (\ref{1.8}), the dynamical evolution of the spinors $\pi^{A},\eta^{A}$ 
(omitting for brevity the constant $k$)
is given by
\begin{equation} \label{2.1}
\begin{array}{c}
\dot{\pi}^{A}=\phi^{AB}\pi_{B}, \\
{ } \\
\dot{\eta}^{A}=\phi^{AB}\eta_{B}. \\
\end{array}
\end{equation}
Again, in the specific case of an electromagnetic field acting on the
particle, these two relations are fully equivalent to the Lorentz Force.
Note that the evolution for both $\pi^{A}$ and $\eta^{A}$ is given by
the same differential form, so up to constant factors, they have the
same analytic solution. On the other side, by carrying out
\begin{equation} \label{2.3}
\begin{array}{c}
\dot{\left(\pi^{A}\eta_{A}\right)}=\dot{\pi}^{A}\eta_{A}+
\pi^{A}\dot{\eta}_{A} \\
{ } \\
=\phi^{AB}\pi_{B}\eta_{A}+\pi^{A}\phi^{BC}\eta_{C}\epsilon_{BA}=
\phi^{AB}\pi_{B}\eta_{A}-\pi_{B}\phi^{BC}\eta_{C} \\
{ } \\
=\phi^{AB}\pi_{B}\eta_{A}-\phi^{BC}\pi_{B}\eta_{C}= \phi^{AB}
\pi_{B}\eta_{A}-\phi^{CB}\pi_{B}\eta_{C}= 0 \\
\end{array}
\end{equation}
we see that $\pi^{A}\eta_{A}$ is a constant of motion in turn related,
as we have already seen, to the mass of the particle by (\ref{1.7}). This
result has been already put forward by A. Bette and J. Buitrago 
\cite{Congress}. Given the components of $\pi^{A},\dot{\pi}^{A}$, referred
to the $spin$ $basis$ $\left\{o^{A}=(0,1),i^{A}=(1,0)\right\}$ as
\begin{displaymath}
\pi^{A}=
\left(\begin{array}{c}
\pi^{0} \\
\pi^{1} \\
\end{array}\right), \quad  \dot{\pi}^{A}= 
\left(\begin{array}{c}
\dot{\pi}^{0} \\
\dot{\pi}^{1} \\
\end{array}\right)
\end{displaymath}
we then obtain a set of two-coupled-differential equations, describing
the dynamical evolution of the particle \\
\begin{displaymath}
\begin{array}{c}
\dot{\pi}^{A}o_{A}=\dot{\pi}^{1}=\phi^{AB}o_{A}\pi_{B}=
\phi^{0B}o_{0}\pi^{B}+ \phi^{1B}o_{1}\pi^{B}=\phi^{1B}\pi_{B}, \\
{ } \\
\dot{\pi}^{A}i_{A}=-\dot{\pi}^{0}=\phi^{AB}i_{A}\pi_{B}=
\phi^{0B}i_{0}\pi^{B}+ \phi^{1B}i_{1}\pi^{B}=-\phi^{0B}\pi_{B}, \\
{ } \\
\dot{\pi}^{1}=\phi^{10}\pi_{0}+\phi^{11}\pi_{1}=-\phi^{10}\pi^{1}+
\phi^{11}\pi^{0}, \\
{ } \\
\dot{\pi}^{0}=\phi^{00}\pi_{0}+\phi^{01}\pi_{1}=-\phi^{00}\pi^{1}+
\phi^{01}\pi^{0}, \\
\end{array}
\end{displaymath} \\
Consider now a four-vector spinorially represented by
\begin{displaymath}
k^{AA'}=\left( \alpha\pi^{A}+\beta\eta^{A}\right)\cdot\left(
\gamma\bar{\pi}^{A'}+\tau\bar{\eta}^{A'}\right)
\end{displaymath}
with coefficients $\alpha,\beta,\gamma,\tau$ subject to the conditions
\begin{displaymath}
\begin{array}{c}
\alpha\gamma=\overline{\alpha\gamma},
{ } \\
\beta\tau=\overline{\beta\tau},
{ }\\
\alpha\tau=\overline{\alpha\tau},
\end{array}
\end{displaymath}
whenever $k^{AA'}$ is to represent a physical four-vector. Then with
the equation of motion (\ref{1.10}) it is easy to show
that the evolution of $k^{AA'}$ obeys to
\begin{equation} \label{2.4}
\dot{k}^{AA'}= F^{AA'BB'}k_{BB'}
\end{equation}
and, in particular, this holds for the following three space-like,
dimensionless \footnote{Note that in natural units the angular momentum
is also dimensionless.}
four-vectors spinorially defined \cite{Congress} by
\begin{equation} \label{2.5}
s^{AA'}=\frac{\mbox{\normalsize{$\pi^{A}\bar{\pi}^{A'}-
\eta^{A}\bar{\eta}^{A'}$}}}{\mbox{\normalsize{$m$}}} ,
\end{equation}
\begin{equation} \label{2.6}
v^{AA'}=\frac{\mbox{\normalsize{$\pi^{A}\bar{\eta}^{A'}+
\eta^{A}\bar{\pi}^{A'}$}}}{\mbox{\normalsize{$m$}}} ,
\end{equation}
\begin{equation} \label{2.7}
w^{AA'}=\frac{\mbox{\normalsize{$i\left(\pi^{A}\bar{\eta}^{A'}-
\eta^{A}\bar{\pi}^{A'}\right)$}}}{\mbox{\normalsize{$m$}}} ,
\end{equation}
which fullfill the conditions
\begin{equation} \label{2.8}
s^{a}s_{a}=v^{a}v_{a}=w^{a}w_{a}=-1,
\end{equation}
{ } \\
\begin{equation} \label{2.9}
s^{a}p_{a}=s^{a}v_{a}=s^{a}w_{a}=0,
\end{equation}
{ } \\
\begin{equation} \label{2.10}
v^{a}p_{a}=v^{a}w_{a}=0, \\
w^{a}p_{a}=0. \\
\end{equation}
and constitute, together with $p^{AA'}$ an orthogonal tetrad.
The ordinary physical components of these vectors are obtained
in the usual way by
\begin{equation} \label{2.11}
s^{AA'}i_{A}\bar{i}_{A'}=\frac{\mbox{\normalsize{$s^{0}+s^{3}$}}}
{\mbox{\normalsize{$\sqrt{2}$}}}, \quad
s^{AA'}o_{A}\bar{o}_{A'}=\frac{\mbox{\normalsize{$s^{0}-s^{3}$}}}
{\mbox{\normalsize{$\sqrt{2}$}}}
\end{equation}
{ } \\
\begin{equation} \label{2.12}
s^{AA'}o_{A}\bar{i}_{A'}=\frac{\mbox{\normalsize{$s^{1}+is^{2}$}}}
{\mbox{\normalsize{$\sqrt{2}$}}}, \quad
s^{AA'}\bar{o}_{A'}i_{A}=\frac{\mbox{\normalsize{$s^{1}-is^{2}$}}}
{\mbox{\normalsize{$\sqrt{2}$}}}
\end{equation}
{ } \\
\begin{equation} \label{2.13}
v^{AA'}i_{A}\bar{i}_{A'}=\frac{\mbox{\normalsize{$v^{0}+v^{3}$}}}
{\mbox{\normalsize{$\sqrt{2}$}}}, \quad
v^{AA'}o_{A}\bar{o}_{A'}=\frac{\mbox{\normalsize{$v^{0}-v^{3}$}}}
{\mbox{\normalsize{$\sqrt{2}$}}}
\end{equation}
{ } \\
\begin{equation} \label{2.14}
v^{AA'}o_{A}\bar{i}_{A'}=\frac{\mbox{\normalsize{$v^{1}+iv^{2}$}}}
{\mbox{\normalsize{$\sqrt{2}$}}}, \quad
v^{AA'}\bar{o}_{A'}i_{A}=\frac{\mbox{\normalsize{$v^{1}-iv^{2}$}}}
{\mbox{\normalsize{$\sqrt{2}$}}}
\end{equation}
{ } \\
\begin{equation} \label{2.15}
w^{AA'}i_{A}\bar{i}_{A'}=\frac{\mbox{\normalsize{$w^{0}+w^{3}$}}}
{\mbox{\normalsize{$\sqrt{2}$}}}, \quad
w^{AA'}o_{A}\bar{o}_{A'}=\frac{\mbox{\normalsize{$w^{0}-w^{3}$}}}
{\mbox{\normalsize{$\sqrt{2}$}}}
\end{equation}
{ } \\
\begin{equation} \label{2.16}
w^{AA'}o_{A}\bar{i}_{A'}=\frac{\mbox{\normalsize{$w^{1}+iw^{2}$}}}
{\mbox{\normalsize{$\sqrt{2}$}}}, \quad
w^{AA'}\bar{o}_{A'}i_{A}=\frac{\mbox{\normalsize{$w^{1}-iw^{2}$}}}
{\mbox{\normalsize{$\sqrt{2}$}}}
\end{equation} \\
In the particle's (instantaneous) proper frame, these four-vectors
become purely spacial  (they will be further identified 
with the components of an intrinsic angular momentum)
and so any linear combination $J^{\alpha}$ of them,
which should evolve at any arbitrary frame, according to
\begin{displaymath}
\dot{J}_{\alpha}=F_{\alpha\beta}J^{\beta}.
\end{displaymath}
%
%
% SECTION INTRINSIC DEGREES OF FREEDOM OF MASSIVE AND MASSLESS PARTICLES
%
%
%
\section{INTRINSIC DEGREES OF FREEDOM OF MASSIVE AND MASSLESS PARTICLES}
So far, we have seen how the requirement within this formalism of expressing
the four-momentum of a massive particle as superposition of two
null-directions, as
\begin{displaymath}
p^{AA'}=\pi^{A}\bar{\pi}^{A'}+\eta^{A}\bar{\eta}^{A'}
\end{displaymath}
lead us, in a natural way, to define the set of three
dynamical additional four-vectors
\begin{displaymath} 
s^{AA'}=\frac{\mbox{\normalsize{$\pi^{A}\bar{\pi}^{A'}-
\eta^{A}\bar{\eta}^{A'}$}}}{\mbox{\normalsize{$m$}}} ,
\end{displaymath}
\begin{displaymath} 
v^{AA'}=\frac{\mbox{\normalsize{$\pi^{A}\bar{\eta}^{A'}+
\eta^{A}\bar{\pi}^{A'}$}}}{\mbox{\normalsize{$m$}}} ,
\end{displaymath}
\begin{displaymath} 
w^{AA'}=\frac{\mbox{\normalsize{$i\left(\pi^{A}\bar{\eta}^{A'}-
\eta^{A}\bar{\pi}^{A'}\right)$}}}{\mbox{\normalsize{$m$}}} ,
\end{displaymath}
which, together with $p^{AA'}$, form a tetrad that appears to be something
inherent to the particle under consideration, merely as a consequence 
of following a trajectory in space-time. In the previous $Section$
we have already  pointed out that they may be associated to internal
degrees of freedom and in the next one, we shall see that they do behave like 
components of an intrinsic angular momentum. Here
we shall try to get some general information about them by analyzing
their components in the (instantaneous) rest frame, for intrinsic properties
such as the spin make sense only in this inertial frame. 
Throughout this section, and just
for brief notation, any quantity is understood to be referred to
the particle's proper frame, in which the conditions
\begin{displaymath}
\begin{array}{c}
\vec{p}=\vec{0},\\
{ } \\
s^{0}=v^{0}=w^{0}=0. \\
\end{array}
\end{displaymath}
must hold. By expressing $\pi^{A},\eta^{A}$ in a general way as
\begin{equation} \label{3.1}
\pi^{A}=
\left(\begin{array}{c}
\vert \pi^{0} \vert e^{i\phi_{0}} \\
\vert \pi^{1} \vert e^{i\phi_{1}} \\
\end{array}\right),
\eta^{A}=
\left(\begin{array}{c}
\vert \eta^{0} \vert e^{i\xi_{0}} \\
\vert \eta^{1} \vert e^{i\xi_{1}} \\
\end{array}\right),
\end{equation}
and from the conditions
\begin{equation} \label{3.2}
p_{3}=0 \Longrightarrow \vert \pi^{0} \vert^{2} +
\vert \eta^{0} \vert^{2}  - \vert \pi^{1} \vert^{2} -
\vert \eta^{1} \vert^{2} = 0,
\end{equation}
\begin{equation} \label{3.3}
s_{0}=0 \Longrightarrow \vert \pi^{0} \vert^{2} -
\vert \eta^{0} \vert^{2}  + \vert \pi^{1} \vert^{2} -
\vert \eta^{1} \vert^{2} = 0,
\end{equation}
we get
\begin{equation} \label{3.4}
\vert \eta^{0} \vert^{2} = \vert \pi^{1} \vert^{2},
\end{equation}
\begin{equation} \label{3.5}
\vert \eta^{0} \vert^{2} = \vert \pi^{1} \vert^{2},
\end{equation}
so $\eta^{A}$ adopts, in this particular frame, the form:
\begin{equation} \label{3.6}
\eta^{A}=
\left(\begin{array}{c}
\vert \pi^{1} \vert e^{i\xi_{0}} \\
\vert \pi^{0} \vert e^{i\xi_{1}} \\
\end{array}\right),
\end{equation}
and the components $v^{00'},v^{11'},w^{11'}$ are then given by
\begin{equation} \label{3.7}
\left\lbrace \begin{array}{c}
v^{00'}=\frac{\mbox{\normalsize{$2$}}}{\mbox{\normalsize{$m$}}}
\vert \pi^{0} \pi^{1} \vert \cos (\phi_{0}-\xi_{0} ) \\
{ } \\
v^{11'}=\frac{\mbox{\normalsize{$2$}}}{\mbox{\normalsize{$m$}}}
\vert \pi^{0} \pi^{1} \vert \cos (\phi_{1}-\xi_{1} ) \\
\end{array}\right\rbrace
\Longrightarrow
\cos(\phi_{0}-\xi_{0})+\cos(\phi_{1}-\xi_{1})=0,
\end{equation}
{ } \\
\begin{equation} \label{3.8}
\left\lbrace \begin{array}{c}
w^{00'}=-\frac{\mbox{\normalsize{$2$}}}{\mbox{\normalsize{$m$}}}
\vert \pi^{0} \pi^{1} \vert \sin (\phi_{0}-\xi_{0} ) \\
{ } \\
w^{11'}=-\frac{\mbox{\normalsize{$2$}}}{\mbox{\normalsize{$m$}}}
\vert \pi^{0} \pi^{1} \vert \cos (\phi_{1}-\xi_{1} ) \\
\end{array}\right\rbrace
\Longrightarrow
\sin(\phi_{0}-\xi_{0})+\cos(\phi_{1}-\xi_{1})=0.
\end{equation}
The conditions given in (\ref{3.7}),(\ref{3.8}) are necessary in order
to guarantee that $v^{0}=w^{0}=0$, so we find that the phase factors
$\phi_{0},\phi_{1},\xi_{0},\xi_{1}$ must be related by
\begin{equation} \label{3.9}
\left( \phi_{1} - \xi_{1} \right) =
\left( \phi_{0} - \xi_{0} \right) + \left( 2n+1 \right) \pi
\end{equation}
or equivalently
\begin{displaymath}
\left( \phi_{1} - \phi_{0} \right) =
\left( \xi_{1} - \xi_{0} \right) + \left( 2n+1 \right) \pi .
\end{displaymath}
This, together with the conditions given in (\ref{3.4}) and
(\ref{3.5}), leads to the automatic fullfillment of 
\begin{displaymath}
\vec{s} \cdot \vec{v} =
\vec{s} \cdot \vec{w} =
\vec{v} \cdot \vec{w} = 0.
\end{displaymath}
On the other side, in the particle's proper frame, the two null-directions
$\pi^{A}\bar{\pi}^{A'},\eta^{A}\bar{\eta}^{A'}$ must appear to have
opposite momentum in order to satisfy the condition above mentioned
($\vec{p}=\vec{0}$). Therefore, the energy associated with each of them
must be the same and subsequently equal to $m/2$, a result that may be
checked through the relations
\begin{equation} \label{3.10}
\left\lbrace \begin{array}{c}
2p^{0}_{\pi}=\sqrt{2} \left[ \vert \pi^{0} \vert^{2} +
\vert \pi^{1} \vert^{2} \right] \\
{ } \\
2p^{0}_{\eta}=\sqrt{2} \left[ \vert \pi^{0} \vert^{2} +
\vert \pi^{1} \vert^{2} \right] \\
\end{array}\right\rbrace
\Longrightarrow
p^{0}_{\pi}=p^{0}_{\eta}=\omega,
\end{equation}
{ } \\
\begin{equation} \label{3.11}
\left\lbrace \begin{array}{c}
2p^{3}_{\pi}=\sqrt{2} \left[ \vert \pi^{0} \vert^{2} -
\vert \pi^{1} \vert^{2} \right] \\
{ } \\
2p^{3}_{\eta}=\sqrt{2} \left[ \vert \pi^{1} \vert^{2} -
\vert \pi^{0} \vert^{2} \right] \\
\end{array}\right\rbrace
\Longrightarrow
p^{3}_{\pi}=-p^{3}_{\eta},
\end{equation}
{ } \\
\begin{equation} \label{3.12}
\left\lbrace \begin{array}{c}
p^{1}_{\pi}=\sqrt{2} \vert \pi^{0}\pi^{1} \vert \cos(\phi_{0}-\phi_{1}) \\
{  } \\
p^{1}_{\eta}=\sqrt{2} \vert \pi^{0}\pi^{1} \vert \cos(\xi_{0}-\xi_{1}) \\
\end{array}\right\rbrace
\Longrightarrow
p^{1}_{\pi}=-p^{1}_{\eta},
\end{equation}
{ } \\
\begin{equation} \label{3.13}
\left\lbrace \begin{array}{c}
p^{2}_{\pi}=\sqrt{2} \vert \pi^{0}\pi^{1} \vert \sin(\phi_{0}-\phi_{1}) \\
{  } \\
p^{2}_{\eta}=\sqrt{2} \vert \pi^{0}\pi^{1} \vert \sin(\xi_{0}-\xi_{1}) \\
\end{array}\right\rbrace
\Longrightarrow
p^{2}_{\pi}=-p^{2}_{\eta},
\end{equation}\\
{ } \\
where $\omega$ is related to the rest-energy of the particle by
\begin{displaymath}
2\omega=m .
\end{displaymath}
Now, in terms of $\omega$ the dimensionless,
normalized four-vectors $s^{AA'},v^{AA'},w^{AA'}$ are
re-written as \\
\begin{equation} \label{3.14}
\begin{array}{c}
s^{AA'}=\frac{\mbox{\normalsize{$\pi^{A}\bar{\pi}^{A'}-
\eta^{A}\bar{\eta}^{A'}$}}}{\mbox{\normalsize{$2\omega$}}},
{ } \\
{ } \\
v^{AA'}=\frac{\mbox{\normalsize{$\pi^{A}\bar{\eta}^{A'}+
\eta^{A}\bar{\pi}^{A'}$}}}{\mbox{\normalsize{$2\omega$}}},
{ } \\
{ } \\
w^{AA'}=i\frac{\mbox{\normalsize{$\left(\pi^{A}\bar{\eta}^{A'}-
\eta^{A}\bar{\pi}^{A'}\right)$}}}{\mbox{\normalsize{$2\omega$}}},
\end{array}
\end{equation}
Note now that both, $v^{AA'}$ and $w^{AA'}$,
are Lorentz-orthogonal to simultaneously $\pi^{A}\bar{\pi}^{A'},\eta^{A}
\bar{\eta}^{A'}$, which spinorially represent null-directions
(e.g. photons) and so the relations
\begin{equation} \label{3.15}
\begin{array}{c}
k^{\mu}k_{\mu}=0,  \\
k^{\mu}v_{\mu}=k^{\mu}w_{\mu}=0, \\
v^{\mu}w_{\mu}=0, \\
\end{array}
\end{equation}
where $k^{\mu}$ represents the four-momentum either of $\pi^{A}
\bar{\pi}^{A'}$ or $\eta^{A}\bar{\eta}^{A'}$, hold. The four-vectors
$v^{\alpha},w^{\alpha}$  may be then identified with polarization vectors
for the null-direction $k^{\mu}$ in the Lorentz-gauge, for which the
condition imposed is  expressed as
\begin{displaymath}
k^{\mu}\epsilon_{\mu}(\vec{k},\lambda)=0.
\end{displaymath}
This allows, as we know, a set of two linearly-independent such vectors
$(\lambda=1,2)$. In fact, for any null-direction spinorially represented
as $\pi^{A}\bar{\pi}^{A'}$, any set of two four-vectors of the form
\begin{displaymath}
\begin{array}{c}
v^{AA'} \propto \left[ \pi^{A}\bar{\xi}^{A'} + \xi^{A}\bar{\pi}^{A'}
\right], \\
w^{AA'} \propto i\left[ \pi^{A}\bar{\xi}^{A'} - \xi^{A}\bar{\pi}^{A'}
\right], \\
\end{array}
\end{displaymath}
satisfies the relations (\ref{3.15}) for any spinor $\xi^{A}$ non-
proportional to $\pi^{A}$. Now, by choosing $\xi^{A}$ as
the counterpart of $\pi^{A}$ giving rise to the null direction $\xi^{A}
\bar{\xi}^{A'}$ with opposite spacial momentum to $\pi^{A}
\bar{\pi}^{A'}$, we have that
\begin{displaymath}
\vec{\epsilon}(\vec{k},\lambda) \cdot \vec{k} =0.
\end{displaymath}
This is then fully equivalent to choose the well-known
$gauge$ $transformation$ $of$ $second$ $kind$ known as the $Coulomb$
$gauge$.
This is precisely the behaviour found in the rest frame with respect to
$\eta^{A}$,
as we saw in the relations
(\ref{3.11})-(\ref{3.13}).
In this frame the null directions $\pi^{A}\bar{\pi}^{A'},\eta^{A}
\bar{\eta}^{A'}$ together with the associated four-vectors $v^{AA'},
w^{AA'}$ 
do therefore represent null-mass particles of spin $1$ and energy
$\omega=m/2$.
Finally, the four-vector
$s^{AA'}$  becomes parallel
to $\vec{p}_{\pi}$, as it is shown in the relations  \\
\begin{equation} \label{3.16}
\left\lbrace \begin{array}{c}
s^{00'}= \frac{\mbox{\normalsize{ $\vert \pi^{0} \vert^{2} -
\vert \pi^{1} \vert^{2}$}}}{\mbox{\normalsize{$2\omega$}}}
{ } \\
{ } \\
s^{11'}= \frac{\mbox{\normalsize{ $\vert \pi^{1} \vert^{2} -
\vert \pi^{0} \vert^{2}$}}}{\mbox{\normalsize{$2\omega$}}}
\end{array}\right\rbrace
\Longrightarrow
s^{00'}=0,s^{3}=\frac{\mbox{\normalsize{$1$}}}
{\mbox{\normalsize{$\sqrt{2}$}}}
\frac{\mbox{\normalsize{$\left[\vert\pi^{0}\vert^{2}-
\vert\pi^{1}\vert^{2}\right]$}}}
{\mbox{\normalsize{$\omega$}}} ,
\end{equation}
{ } \\
\begin{equation} \label{3.17}
s^{01'}=\frac{\mbox{\normalsize{$\vert\pi^{0}\pi^{1}\vert$}}}
{\mbox{\normalsize{$2\omega$}}}
\left( e^{i(\phi_{0}-\phi_{1})} - e^{i(\xi_{0}-\xi_{1})}
\right) \Longrightarrow
\left\{ \begin{array}{c}
s^{1}=\frac{\mbox{\normalsize{$1$}}}
{\mbox{\normalsize{$\sqrt{2}$}}}
{\mbox{\normalsize{$\vert\pi^{0}\pi^{1}\vert\cos(\phi_{0}-\phi_{1})$}}}
{\mbox{\normalsize{$\omega$}}} \\
{ } \\
s^{2}=\frac{\mbox{\normalsize{$1$}}}
{\mbox{\normalsize{$\sqrt{2}$}}}
{\mbox{\normalsize{$\vert\pi^{0}\pi^{1}\vert\sin(\phi_{0}-\phi_{1})$}}}
{\mbox{\normalsize{$\omega$}}} \\
\end{array} \right\rbrace .
\end{equation}
%
%
%
% SECTION EVOLUTION OF INTRINSIC SPIN UNDER AN HOMOGENEOUS CONSTANT MAGNETIC 
FIELD
\section{EVOLUTION OF INTRINSIC SPIN UNDER AN HOMOGENEOUS CONSTANT MAGNETIC 
FIELD}
Suppose a massive ($m=1$), charged particle
in the presence of and external, homogeneous, constant magnetic field
directed along the $z$-axis. By applying the stardard Lorentz Force equation
and solving the first order coupled system for the evolution of the
momentum we get
\begin{equation} \label{4.1}
p_{z}(\tau)=constant,
\end{equation}
\begin{equation} \label{4.2}
p_{x}(\tau)=A_{1}\cos(qB\tau)+A_{2}\sin(qB\tau),
\end{equation}
\begin{equation} \label{4.3}
p_{y}(\tau)=A_{2}\cos(qB\tau)-A_{1}\sin(qB\tau),
\end{equation}
where $A_{1}$ and $A_{2}$ are to be determined from the initial conditions.
In this case, the equations to solve are 
\begin{displaymath}
\begin{array}{c}
\dot{\pi}^{0}=-i \frac{\mbox{\normalsize{$qB$}}}{\mbox{\normalsize{$2$}}}
\pi^{0}, \\
{ } \\
\dot{\pi}^{1}=i \frac{\mbox{\normalsize{$qB$}}}{\mbox{\normalsize{$2$}}}
\pi^{1}, \\
\end{array}
\end{displaymath}
(the same holds for $\dot{\eta}^{A}$). Note that, in this case, the equations 
become simpler than in the traditional formulation and will also give the 
evolution of the intrinsic spin. This follows from the fact that, according to 
the geometrical interpretation of the Lorentz Force \cite{buitrago}, the $\vec 
B$ components are related to rotations; hence to the $SU(2)$ subgroup of 
$SL(2,C)$ related to spin.  The solution to these equations is given by
\begin{displaymath}
\begin{array}{c}
\pi^{0}(\tau)=\pi^{0}(\tau=0)exp \left(
-i \frac{\mbox{\normalsize{$qB$}}}{\mbox{\normalsize{$2$}}} \tau \right),
{ } \\
{ } \\
\pi^{1}(\tau)=\pi^{1}(\tau=0)exp \left(
i \frac{\mbox{\normalsize{$qB$}}}{\mbox{\normalsize{$2$}}} \tau \right),
{ } \\
{ } \\
\eta^{0}(\tau)=\eta^{0}(\tau=0)exp \left(
-i \frac{\mbox{\normalsize{$qB$}}}{\mbox{\normalsize{$2$}}} \tau \right),
{ } \\
{ } \\
\eta^{1}(\tau)=\eta^{1}(\tau=0)exp \left(
i \frac{\mbox{\normalsize{$qB$}}}{\mbox{\normalsize{$2$}}} \tau \right),
\end{array}
\end{displaymath}
and therefore \\
\begin{displaymath}
\begin{array}{c}
p^{00'}=\vert \pi^{0}(\tau=0)\vert^{2}+ \vert \eta^{0}(\tau=0)\vert^{2}=
constant, \\
{ } \\
p^{01'}=\pi^{0}(\tau=0)\bar{\pi}^{1'}(\tau=0) exp \left(-iqB\tau\right)+
\eta^{0}(\tau=0)\bar{\eta}^{1'}(\tau=0) exp \left(-iqB\tau\right)=
A(\tau=0)exp \left(-iqB\tau \right), \\
{ } \\
p^{11'}=\vert \pi^{1}(\tau=0)\vert^{2}+ \vert \eta^{1}(\tau=0)\vert^{2}=
constant, \\
\end{array}
\end{displaymath}
{ } \\
where $A(\tau=0)$ is a complex number whose value is to be determined from
the initial conditions, so $p^{01'}$ can be written in the form
{ } \\
\begin{equation} \label{4.4}
p^{01'}(\tau)=A_{1}\cos(qB\tau)+A_{2}\sin(qB\tau)+
i\left[ A_{2}\cos(qB\tau)-A_{1}\sin(qB\tau)\right]
\end{equation}
{ } \\
being $A_{1}=Re\{A(\tau=0)\}$ and $A_{2}=Im\{A(\tau=0)\}$. We get then
for the components of the four-momentum:
\begin{equation} \label{4.5}
\left\{ \begin{array}{c}
\sqrt{2}p^{00'}= E+p_{z}=constant \\
\sqrt{2}p^{11'}= E-p_{z}=constant \\
\end{array}\right\} \Longrightarrow
E,p_{z}=constant,
\end{equation}
{ } \\
\begin{equation} \label{4.6}
\sqrt{2}p^{01'}=p_{x}+ip_{y} \Longrightarrow
\left\{ \begin{array}{c}
p_{x}=\sqrt{2} \left[A_{1}\cos(qB\tau)+A_{2}\sin(qB\tau) \right]
{ } \\
p_{y}=\sqrt{2} \left[A_{2}\cos(qB\tau)-A_{1}\sin(qB\tau)\right]
\end{array} \right\}
\end{equation}
which, as expected, coincides with the well-known solutions. For the
time evolution of the four-vectors $s^{a},v^{a},w^{a}$ we find
\begin{equation} \label{4.7}
\left\{ \begin{array}{c}
\sqrt{2}s^{00'}=\frac{\mbox{\normalsize{$\sqrt{2}$}}}{\mbox{\normalsize{
$m$}}} \left[ \vert \pi^{0}(\tau=0) \vert^{2} -
\vert \eta^{0}(\tau=0) \vert^{2} \right] =
s^{0}+ s^{z}
{ } \\
\sqrt{2}s^{11'}=\frac{\mbox{\normalsize{$\sqrt{2}$}}}{\mbox{\normalsize{
$m$}}} \left[ \vert \pi^{1}(\tau=0) \vert^{2} -
\vert \eta^{1}(\tau=0) \vert^{2} \right] =
s^{0}- s^{z}
\end{array}\right\} \Longrightarrow
s^{0},s^{z}=constant,
\end{equation}
{ } \\
\begin{equation} \label{4.8}
\left\{ \begin{array}{c}
\sqrt{2}v^{00'}=\frac{\mbox{\normalsize{$\sqrt{2}$}}}{\mbox{\normalsize{
$m$}}} \left[ \pi^{0}(\tau=0)\bar{\eta}^{0}(\tau=0)+c.c.\right] =
v^{0}+ v^{z}
{ } \\
\sqrt{2}v^{11'}=\frac{\mbox{\normalsize{$\sqrt{2}$}}}{\mbox{\normalsize{
$m$}}} \left[ \pi^{1}(\tau=0)\bar{\eta}^{1}(\tau=0) +c.c. \right] =
v^{0}- v^{z}
\end{array}\right\} \Longrightarrow
v^{0},v^{z}=constant,
\end{equation}
{ } \\
\begin{equation} \label{4.9}
\left\{ \begin{array}{c}
\sqrt{2}w^{00'}=\frac{\mbox{\normalsize{$\sqrt{2}$}}}{\mbox{\normalsize{
$m$}}} \left[ i\pi^{0}(\tau=0)\bar{\eta}^{0}(\tau=0)+c.c.\right] =
w^{0}+ w^{z}
{ } \\
\sqrt{2}w^{11'}=\frac{\mbox{\normalsize{$\sqrt{2}$}}}{\mbox{\normalsize{
$m$}}} \left[ i\pi^{1}(\tau=0)\bar{\eta}^{1}(\tau=0) +c.c. \right] =
w^{0}- w^{z}
\end{array}\right\} \Longrightarrow
w^{0},w^{z}=constant,
\end{equation}
{ } \\
\begin{equation} \label{4.10}
\sqrt{2}s^{01'}=s_{x}+is_{y}=B(\tau=0)e^{(-qB\tau)} \Longrightarrow
\left\{ \begin{array}{c}
s_{x}=B_{1}\cos(qB\tau)+B_{2}\sin(qB\tau) \\
{ } \\
s_{y}=B_{2}\cos(qB\tau)-B_{1}\sin(qB\tau) \\
\end{array} \right\} ,
\end{equation}
{ } \\
\begin{equation} \label{4.11}
\sqrt{2}v^{01'}=v_{x}+iv_{y}=C(\tau=0)e^{(-qB\tau)} \Longrightarrow
\left\{ \begin{array}{c}
v_{x}=C_{1}\cos(qB\tau)+C_{2}\sin(qB\tau) \\
{ } \\
v_{y}=C_{2}\cos(qB\tau)-C_{1}\sin(qB\tau) \\
\end{array} \right\},
\end{equation}
{ } \\
\begin{equation} \label{4.12}
\sqrt{2}w^{01'}=w_{x}+iw_{y}=D(\tau=0)e^{(-qB\tau)} \Longrightarrow
\left\{ \begin{array}{c}
w_{x}=D_{1}\cos(qB\tau)+D_{2}\sin(qB\tau) \\
{ } \\
w_{y}=D_{2}\cos(qB\tau)-D_{1}\sin(qB\tau) \\
\end{array} \right\} .
\end{equation}
If the initial conditions are such that $p_{x}(\tau=0)=
p_{y}(\tau=0)=0$ the particle keeps on moving with constant velocity along
the $z$-axis and the (orbital) angular momentum is therefore null. However,
if the particle is to have an intrinsic angular momentum, although no
Thomas precession occurs, we should expect its intrinsic magnetic moment
to precess around the $z$-axis, due to the presence of the magnetic field.
This is precisely the behaviour found for $\vec{s},\vec{v},\vec{w}$:
they do precess due solely to the presence of the applied magnetic field
and can be therefore identified with the components of an intrinsic
angular momentum. To analyze the situation in the rest frame of the particle,
we set $\vec{p}=0$. Then, we find that the temporal evolution of the
spinors in this frame is
\begin{equation} \label{4.13}
\pi^{A}(\tau)= \left( \begin{array}{c}
\vert \pi^{0} \vert (\tau=0) e^{i\phi_{0}}e^{i\frac{qB\tau}{2}} \\
\vert \pi^{1} \vert (\tau=0) e^{i\phi_{1}}e^{i\frac{qB\tau}{2}} \\
\end{array} \right),
\eta^{A}(\tau)= \left( \begin{array}{c}
\vert \pi^{1} \vert (\tau=0) e^{i\xi_{0}}e^{i\frac{qB\tau}{2}} \\
\vert \pi^{0} \vert (\tau=0) e^{i\xi_{1}}e^{i\frac{qB\tau}{2}} \\
\end{array} \right),
\end{equation}
where the phase-factors $\phi_{0},\phi_{1},\xi_{0},\xi_{1}$ are
related by (\ref{3.9}). For the momenta of the null directions associated
to massless particles in this inertial frame we find
\begin{equation} \label{4.14}
\vec{p}_{\pi}=
\left(
\frac{\mbox{\normalsize{$\vert\pi^{0}\pi^{1}\vert$}}}
{\mbox{\normalsize{$\sqrt{2}$}}} \cos(qB\tau+\delta\phi),
\frac{\mbox{\normalsize{$\vert\pi^{0}\pi^{1}\vert$}}}
{\mbox{\normalsize{$\sqrt{2}$}}} \sin(qB\tau+\delta\phi),
\sqrt{2} \left[\vert\pi^{0}\vert^{2}-\vert\pi^{1}\vert^{2}\right]
\right) ,
\end{equation}
\begin{equation} \label{4.15}
\vec{p}_{\eta}=- \left(
\frac{\mbox{\normalsize{$\vert\pi^{0}\pi^{1}\vert$}}}
{\mbox{\normalsize{$\sqrt{2}$}}} \cos(qB\tau+\delta\phi),
\frac{\mbox{\normalsize{$\vert\pi^{0}\pi^{1}\vert$}}}
{\mbox{\normalsize{$\sqrt{2}$}}} \sin(qB\tau+\delta\phi),
\sqrt{2} \left[\vert\pi^{0}\vert^{2}-\vert\pi^{1}\vert^{2}\right]
\right) ,
\end{equation}
with $\delta\phi=(\phi_{0}-\phi_{1})$. \\
We see in (\ref{4.14}) and (\ref{4.15}) that the $z$-component
remains constant in time, while the $x$- and $y$-components do
evolve. This causes, in this particular case, the rotation of  $\vec{p}_{\pi}$ 
and $\vec{p}_{\eta}$ around the
$z$-axis. With use of the equations (\ref{1.13}),(\ref{1.14}),(\ref{1.15})
one obtains (note that $\vec{s}$ is aligned with $\vec{p}_{\pi}$)
\begin{equation} \label{4.16}
\vec{s}= \frac{\mbox{\normalsize{$\vec{p}_{\pi}$}}}
{\mbox{\normalsize{$m$}}},
\end{equation}
\begin{equation} \label{4.17}
 \begin{array}{c}
v_{x}= \frac{\mbox{\normalsize{$\sqrt{2}$}}}
{\mbox{\normalsize{$m$}}} \left\{
\vert\pi^{0}\vert^{2}\cos(qB\tau+\phi_{0}-\xi_{1})-
\vert\pi^{1}\vert^{2}\cos(qB\tau+\xi_{0}-\phi_{1})
\right\}, \\
{ } \\
v_{y}= \frac{\mbox{\normalsize{$\sqrt{2}$}}}
{\mbox{\normalsize{$m$}}} \left\{
\vert\pi^{0}\vert^{2}\sin(qB\tau+\phi_{0}-\xi_{1})-
\vert\pi^{1}\vert^{2}\sin(qB\tau+\xi_{0}-\phi_{1})
\right\}, \\
{ } \\
v_{z}= \frac{\mbox{\normalsize{$2\sqrt{2}$}}}
{\mbox{\normalsize{$m$}}}
\vert\pi^{0}\pi^{1}\vert\cos(\phi_{0}-\xi_{0}), \\
\end{array} 
\end{equation}
\begin{equation} \label{4.18}
 \begin{array}{c}
w_{x}= \frac{\mbox{\normalsize{$\sqrt{2}$}}}
{\mbox{\normalsize{$m$}}} \left\{
\vert\pi^{1}\vert^{2}\sin(qB\tau+\xi_{0}-\phi_{1})-
\vert\pi^{0}\vert^{2}\sin(qB\tau+\phi_{0}-\xi_{1})
\right\}, \\
{ } \\
w_{y}= \frac{\mbox{\normalsize{$\sqrt{2}$}}}
{\mbox{\normalsize{$m$}}} \left\{
\vert\pi^{0}\vert^{2}\cos(qB\tau+\phi_{0}-\xi_{1})-
\vert\pi^{1}\vert^{2}\cos(qB\tau+\xi_{0}-\phi_{1})
\right\}, \\
{ } \\
w_{z}= - \frac{\mbox{\normalsize{$2\sqrt{2}$}}}
{\mbox{\normalsize{$m$}}}
\vert\pi^{0}\pi^{1}\vert\sin(\phi_{0}-\xi_{0}). \\
\end{array} 
\end{equation}
Notice that $\vec{s},\vec{v},\vec{w}$ keep
instantaneously, at any proper time $\tau$, orthogonal to each other
while undergoing the precession around the $z$-axis. This will also happen to 
any vector formed as a linear combination of them. However, this point needs 
further clarification which will be the subject on next section.
%
%Section SPINORIAL REPRESENTATION OF AN INTRINSIC SPIN

\section{SPINORIAL REPRESENTATION OF AN INTRINSIC SPIN}
So far, we have seen how the three four-vectors $s^{AA'},v^{AA'},w^{AA'}$ behave 
in the particle's proper frame as components of an intrinsic angular momentum 
vector under the action of an applied external magnetic field. This fact seems 
to indicate that these four-vectors are, somehow, related to the intrinsic spin 
of the particle. To elucidate further this point, we switch off the magnetic 
field and recall that, in the particle's proper frame, the spinors $\pi^A 
\bar\pi^{A'}$ and $\eta^A \bar\eta^{A'}$ do represent null-directions having the 
same energy and opposite spacial momentum. However, the direction of $\vec 
p_\pi$ is absolutely arbitrary and we can take it, for instance, along the 
$z-direction$. With this we have that

\begin{equation}
\pi^{A}=
\left(\begin{array}{c}
\pi^{0} \\
0 \\
\end{array}\right), \quad  \eta^{A}= 
\left(\begin{array}{c}
0 \\
\eta^{1} \\
\end{array}\right).
\end{equation}

With this representation, $\pi^A \pi^{A'}$ represents a massless particle moving 
along the $z-direction$, in positive sense. Next, the condition of having the 
same energy imposes $\vert \eta^1 \vert = \vert \pi^0 \vert $; with this, the 
more general expression for both $\pi^A, \eta^A$ is

\begin{equation}
\pi^{A}=\vert \pi^0\vert
\left(\begin{array}{c}
e^{i\phi_0} \\
0 \\
\end{array}\right), \quad  \eta^{A}=\vert \pi^0 \vert 
\left(\begin{array}{c}
0 \\
e^{i\xi_1} \\
\end{array}\right).
\end{equation}

Again there is arbitrariness with respect to the phase factors $\phi_0, \xi_1$. 
We shall partially remove it by setting $\xi_1 = \phi_0$. Next, remember that, 
although $s^{AA'},v^{AA'},w^{AA'}$ are four-vectors, they have also an 
associated matrix representation,
in such a way that its determinant coincides with the Lorentz-norm of the
corresponding four-vector. Taking into account the condition (7), these 
associated matrices
are given by

\begin{equation}  \label{Pauli}
s^{AA'}=  
\left(\begin{array}{clrr}
1 &  \quad 0 \\
0 & -1 \\
\end{array}\right), \quad  v^{AA'}=   
\left(\begin{array}{clrr}
0 & \quad 1\\
1 & \quad 0 \\
\end{array}\right), \quad w^{AA'}= 
\left( \begin {array}{clrr}
0 & \quad i\\
-i & \quad 0 \\
\end{array}\right).
\end{equation}
The reader will immediately recognize the three Pauli-spin matrices
 \footnote{Although the usual representation of the Pauli-spin-matrices sets 
 $\epsilon_{123}=1$, due to the global sign chose for $w^{AA'}$, 
 the representation here given corresponds to $\epsilon_{123}=-1$.
 A global sign
 however does not affect previous or further results.},
which represent the intrinsic angular momentum of a one-half spin
massive particle.
Since equation (\ref{Pauli}) correspond to four-vectors with null-time
component, following the steps of Quantum Mechanics, we could define the
$spin$ $operator$ of the particle, at its rest-frame, as
\begin{equation}
\vec{S}=\frac{1}{2}\left(s^{AA'},v^{AA'},w^{AA'}\right)
\end{equation}
and related through a Lorentz transformation to any other inertial frame.
%
% FINAL REMARKS
%
%
%
\section*{FINAL REMARKS}
According to the principle presented in \cite{buitrago} and summarized
in $Section$ $I$, the form of the classical equations of motion can
be regarded as a consequence of the geometry of the complex two-dimensional
$Spin$ $Space$ which sets in turn the local geometry of space-time to
be minkowskian. In this context, spinors are more fundamental
objects than four-vectors and, as seen in IV, V and VI, are able to describe the 
intrinsic spin of massless particles and also the intrinsic spin $1/2$ of 
massive ones . The before mentioned principle lead us to a set of new spinorial 
differential 
equations describing the dynamics of a massive spinning object,
since not only the evolution of the world trajectories are contained
in these equations, but are also susceptible of describing new degrees of 
freedom associated to its intrinsic spin which turns out to be the corresponding 
to spin one-half particles. Finally we would like to point out that the 
space-time dynamics described in this way are performed entirely in spinor space 
and do not make explicit use of the space-time coordinates that become secondary 
constructions. In this sense, our study differ from the twistor approach 
(especially in its two-twistor version, see e.g. \cite{Bette et al.}) which make 
use of the space-time coordinates.
%
%
%
% ACKNOWLEDGMENT
%
%
%
\section*{Acknowledgment}
We would like specially thank Doctor Andreas Bette (deceased in May 2005)
of the Royal Institute of Technology (Sweden), for inspiring some of the
ideas here developed.

\end{document}